%% file: main.tex
\documentclass[runningheads]{llncs}
\usepackage[T1]{fontenc}
\usepackage{graphicx}
\usepackage{tikz}
\usepackage{amsmath}
\usepackage{hyperref}
\usepackage{multirow}
\usetikzlibrary{positioning}
\usetikzlibrary{decorations.text}
\usetikzlibrary{decorations.pathmorphing}
\usetikzlibrary{arrows.meta}
\usepackage[linesnumbered,ruled,vlined]{algorithm2e}
\usepackage{graphicx}
\usepackage{subcaption}
\usepackage{mwe}
\usepackage{xcolor}

\SetCommentSty{mycommfont}

\SetKwInput{KwInput}{Input}                
\SetKwInput{KwOutput}{Output}              

\usetikzlibrary{shapes.callouts}
\tikzset{
  level/.style   = { ultra thick, blue },
  connect/.style = { dashed, red,-> },
  notice/.style  = { draw, rectangle callout, callout relative pointer={#1} },
  label/.style   = { text width=2cm }
}
\tikzset{%
  >={Latex[width=2mm,length=2mm]},
            base/.style = {rectangle, rounded corners, draw=black,
                           minimum width=4cm, minimum height=1cm,
                           text centered, font=\sffamily},
  activityStarts/.style = {base, fill=blue!30},
       startstop/.style = {base, fill=red!30},
    activityRuns/.style = {base, fill=green!30},
         process/.style = {base, minimum width=2.5cm, fill=orange!15,
                           font=\ttfamily},
}
\usetikzlibrary{arrows,automata,chains,matrix,positioning,scopes}
\makeatletter
\tikzset{join/.code=\tikzset{after node path={%
\ifx\tikzchainprevious\pgfutil@empty\else(\tikzchainprevious)%
edge[every join]#1(\tikzchaincurrent)\fi}}}
\makeatother
\tikzset{>=stealth',every on chain/.append style={join},
         every join/.style={->}}
\tikzstyle{labeled}=[execute at begin node=$\scriptstyle,
   execute at end node=$]

\newcommand{\rev}{black}

\begin{document}
\title{Non-Functional Testing of Runtime Enforcers in Android}
%
%
\author{
Oliviero Riganelli\inst{1}\orcidID{0000-0003-2120-2894}\and
Daniela Micucci\inst{1}\orcidID{0000-0003-1261-2234}\and
Leonardo Mariani\inst{1}\orcidID{0000-0001-9527-7042}}
\authorrunning{Riganelli et al.}
%
\institute{University of Milano-Bicocca, Milan 20126, Italy\\
\email{\{oliviero.riganelli,daniela.micucci,leonardo.mariani\}@unimib.it}}

\maketitle              
\begin{abstract}
Runtime enforcers can be used to ensure that running applications satisfy desired correctness properties. Although runtime enforcers that are correct-by-construction with respect to abstract behavioral models are relatively easy to specify, the concrete software enforcers generated from these specifications may easily introduce issues in the target application. Indeed developers can generate test suites to verify the \emph{functional} behavior of the enforcers, for instance exploiting the same models used to specify them. However, it remains challenging and tedious to verify the behavior of enforcers in terms of \emph{non-functional} performance characteristics. This paper describes a practical approach to reveal runtime enforcers that may introduce \emph{inefficiencies} in the target application. The approach relies on a combination of automatic test generation and runtime monitoring of multiple key performance indicators. 
We designed our approach to reveal issues in four indicators for mobile systems: responsiveness, launch time, memory, and energy consumption. Experimental results show that our approach can detect performance issues that might be introduced by automatically generated enforcers.

\end{abstract}
%
%

\input{Introduction}

\input{background}
\input{approach}

\input{evaluation}

\input{RW}
\section{Conclusion and Future Work} \label{sec:conclusion}
Software enforcers can be effectively used to modify executions to ultimately guarantee that correctness policies are satisfied. Since software enforcers are active in the operational environment (e.g., in the end user environment), it is compulsory to use dependable enforcers that cannot affect negatively the target app. It is thus important to extensively validate enforcers before they can be used.

In this paper we presented Test4Enforcers, which can be used to validate the correctness of software enforcers on both a functional and non-functional perspective. Early results show that the generated test cases can feasibly detect problems, preventing the distribution of faulty enforcers. 

Future work mainly concerns with experimenting Test4Enforcers with a larger set of apps and enforcers, and collecting additional evidence about the effectiveness of the approach.

\subsubsection{Acknowledgements} 
We would like to thanks Alice Hoa Galli for her help with the experiments.

%
%
%
\bibliographystyle{splncs04}
\bibliography{main}
\end{document}

%% file: Introduction.tex
\section{Introduction}\label{Introduction}

Mobile applications are extremely popular. Indeed, there are applications to support virtually any task, as witnessed by the more than 3 million applications available for download in Google Play in the first quarter of 2022~\cite{linkStatista01}. 

Mobile applications interact with the hosting device, exploiting the available resources, such as the camera, memory, battery, and Wi-Fi antenna. Unfortunately, mobile applications may easily misuse resources, causing issues to the underlying system and the rest of the applications running in the device. For instance, an application may acquire the camera without releasing it, preventing the access to the camera to the other applications.

To prevent these problems, users of mobile applications can install and activate software enforcers~\cite{Barringer2010,survey2012,Riganelli:ProactiveLibraries:ACMTAAS:2019,Falcone:AndoridEnforcement:RV:2012,DaianFMSSIR15}  that guarantee that specific correctness policies are satisfied (e.g., the camera is always released after it has been acquired). These enforcers can be typically generated automatically from a model-based representation of the processes involved in the policy that must be enforced. For instance, I/O automata can model the behavior of apps and services running in a mobile device, and edit automata~\cite{Ligatti:EditAutomata:2005} can be used to specify enforcers that can \textcolor{\rev}{correct executions to avoid policy violations. Software enforcers derived from these models are guaranteed to fix the execution since they are correct-by-construction, as long as both the specified models and the code generation process are correct.}

However, models are abstractions of the behavior of the software and its environment, and often miss many relevant details that might affect the correctness of the enforcers. For this reason, software enforcers, even when generated automatically, have to be tested. In previous work~\cite{Guzman:Test4Enforcer:RV:2020}, we addressed the challenge of automatically generating test cases that cover the functional specification used to generated the enforcers, and to  \textcolor{\rev}{ensure correctness-by-construction before the enforcers are deployed. However, correctness-by-construction is limited to the functional aspect of the enforcers. In this paper, we address the challenge of enriching the testing strategy with the capability to collect and analyze non-functional indicators to ensure non-functional properties, specifically through the detection of performance problems that might be introduced into the target system in an attempt to apply (functional) policy.}
We experimented our approach with several enforcers, indicators, and faults, demonstrating the usefulness of addressing both the functional and non-functional aspects when verifying software enforcers.  

The paper is organized as follows. Section~\ref{sec:background} provides background information about software enforcement. Section~\ref{sec:approach} describes the test case generation strategy for software enforcers, augmented to deal with performance problems. Section~\ref{sec:evaluation} reports the results that we obtained with the evaluation of the approach. Section~\ref{sec:related} discusses related work. Finally, Section~\ref{sec:conclusion} provides final remarks.

%% file: background.tex
\section{Background }\label{sec:background}
In this section we introduce the notion of runtime policy and policy enforcement. 


\subsection{Runtime Policy}
A runtime policy is a predicate over a set of executions. More formally, let $\Sigma$ be a finite set of observable program actions $a$. An \emph{execution} $\sigma$ is a finite or infinite non-empty sequence of actions $a_1;a_2;\ldots;a_n$. $\Sigma^*$ is the set of all finite sequences, $\Sigma^\omega$ is the set of infinite sequences, and $\Sigma^\infty = \Sigma^* \cup \Sigma^\omega$ is the set of all sequences. Given a set of executions $\chi \subseteq \Sigma^\infty$, a \emph{policy} is a predicate $P$ on $\chi$. A policy $P$ is satisfied by a set of executions $\chi$ if and only if $P(\chi)$ evaluates to $true$.

A policy may concern any behavior of an application, including resource usage and security. For example, an Android policy~\cite{CameraAPI} requires that anytime an Android app stops using the camera, it explicitly releases the camera to make it available to the other apps. More in details, \emph{``if an activity\footnote{Activities are the entry point of interactions between users and apps \href{https://developer.android.com/guide/components/activities}{https://developer.android.com/guide/components/activities}} is using the camera and the activity receives an invocation to the callback method \texttt{onPause()}\footnote{\texttt{onPause()} is a callback method that is invoked by the Android framework every time an activity is paused.}, the activity must release the camera}.'' 
We refer to this policy throughout the paper to describe our approach.

\subsection{Policy Enforcement Models}

\begin{figure}
    \centering
      \resizebox{3in}{!}{%
\begin{tikzpicture}[->,>=stealth',shorten >=1pt,auto,node distance=8.9cm,scale=1,transform shape]

  \node[state,initial] at (-4, 0) (a0p0e0) {$s_0$};

  \node[state] (e2p1a0) [right of=a0p0e0] {$s_1$};

 \path (a0p0e0) edge      [bend left]        node {$camera.open()_{req}?/camera.open()_{api}!$} (e2p1a0)

 (a0p0e0) edge      [loop right = 60]        node {$activity.onPause()_{req}?/activity.onPause()_{api}!$} (a0p0e0)

 (e2p1a0) edge      [bend left=60]        node {$camera.release()_{req}?/camera.release()_{api}!$} (a0p0e0)
 (e2p1a0) edge      [bend right = 60]        node[pos=0.5,above,rotate=0] {$activity.onPause()_{req}?/camera.release()_{api}!;activity.onPause()_{api}!$} (a0p0e0);
\end{tikzpicture}
}%
    \caption{Enforcer for systematically releasing the camera when the activity is paused.}
    \label{fig:enforcer-composition}
\end{figure}
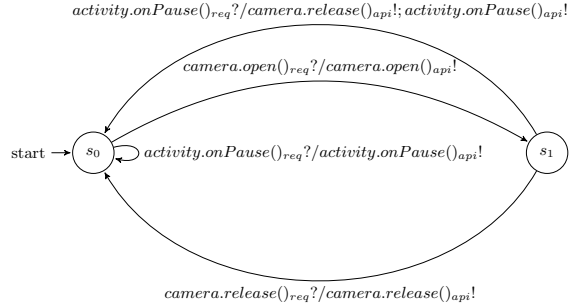

A policy enforcement model specifies how executions must be changed to make them satisfy a given policy. Policy enforcers can be represented with finite-state models, such as edit and input/output automata. For instance, Figure~\ref{fig:enforcer-composition} shows an input/output automaton that specifies an enforcement model that can enforce a faulty activity that does not release the camera to release it.

The inputs are the events intercepted by the enforcer (represented with the \emph{req} subscript in the model)  while the outputs (represented with the \emph{api} subscript) are the events emitted by the enforcer in response to the intercepted events. When the input and the output in a same transition match (regardless of the subscript), the enforcer is not changing the execution. If the output differs from the input, the enforcer is changing the execution suppressing and/or adding events.  

Referring to the example model in Figure~\ref{fig:enforcer-composition}, state $s_0$ represents the case the camera has not been acquired yet, and the enforcer is ready to intercept events without altering the execution. In fact, both pausing the current activity (event \texttt{activity.onPause()}) and acquiring the camera (event \texttt{camera.open()}) are events compatible with the policy related to the access to the camera. Once the camera is acquired (event \texttt{camera.open()} from state $s_0$), the camera must be released before the activity can be paused. In state $s_1$, the enforcer is thus ready to accept the \texttt{camera.release()} event. On the contrary, if the activity is paused, the \texttt{activity.onPause()} event would violate the policy (i.e., the activity would be paused without releasing the camera), and thus the enforcer modifies the execution emitting the sequence of events \texttt{camera.release()} \texttt{activity.onPause()}, which guarantees the satisfaction of the policy.

Enforcement strategies must be translated into software components (i.e., software enforcers) that enforce the specified strategies at runtime. This translation could be done manually, semi-automatically, or automatically. In all the cases, the resulting components may include bugs, due to issues in the translation process and the extra code added by developers to obtain fully functional components, and must be tested extensively before they can be deployed and used. 


To detect functional bugs, we developed Test4Enforcers~\cite{Guzman:Test4Enforcer:RV:2020}, a tool that automatically generates test cases that cover the functional specification used to generate the enforcers. In this paper, we describe how we extended Test4Enforcers with the capability to  detect performance bugs by collecting and analyzing non-functional indicators. 

%% file: approach.tex
\section{Test4Enforcers }\label{sec:approach}
Test4Enforcers detects bugs by implementing a two-step detection strategy.  
The first step, \textit{test case generation}, generates test cases that cover the behavior of the enforcer, considering both the case the enforcer has to modify the execution to enforce the policy and the enforcer does not need to change the execution since it already satisfies the policy. The second step, \textit{test case execution}, executes the generated test cases in the target environment both with and without the policy enforcer by collecting the Key Performance Indicators (KPIs) that are analyzed to detect any non-functional bug.

\subsection{Test Case Generation}
\begin{figure}
    \centering
        \includegraphics[width=\textwidth]{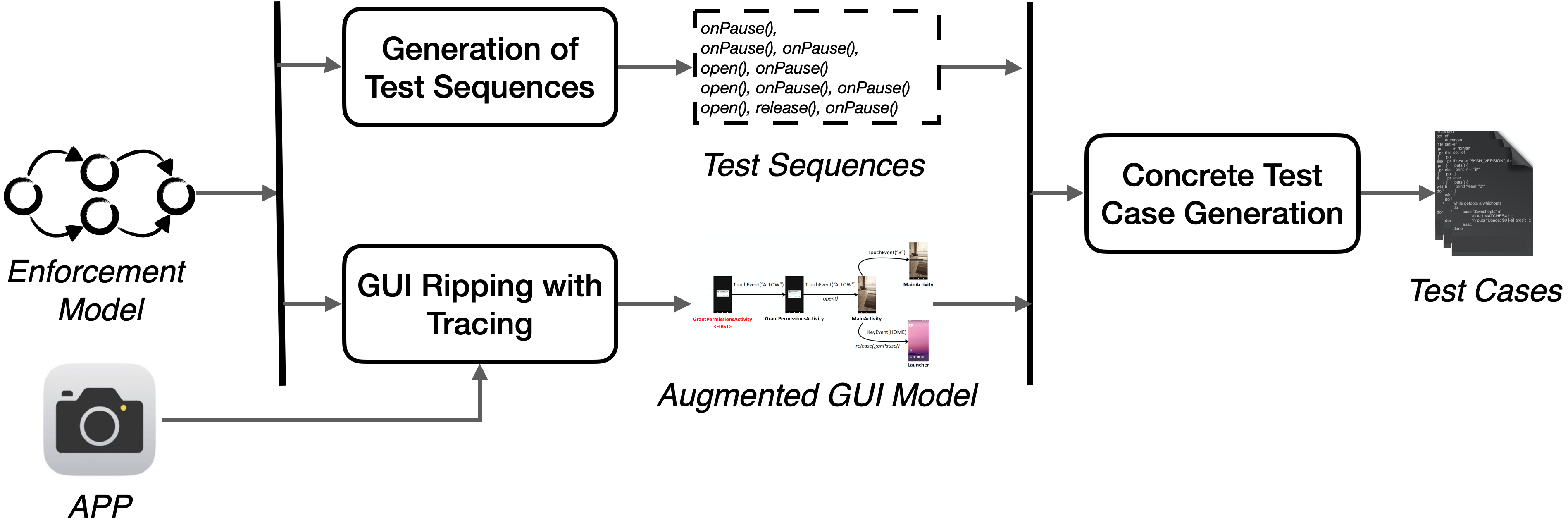}
    \caption{Test case generation with Test4Enforcers.}
    \label{fig:TCGeneration}
\end{figure}

Test case generation consists of 3 activities, as shown in Figure \ref{fig:TCGeneration}. 

\emph{i)} The \textit{Generation of the Test Sequences} activity generates the test sequences that must be covered to thoroughly test the behavior of the enforcer according with the enforcement model.  

\emph{ii)} The \textit{GUI Ripping with Tracing} activity runs a GUI Ripping process that explores the GUI of the app under test while tracing the events that are in the alphabet of the enforcement model. The output of this activity is an augmented GUI model with states representing the GUI states visited during the ripping process and the transitions representing the GUI actions that caused the state change. The model is augmented since each transition is annotated with the events in the alphabet of the enforcement model that have been executed as a consequence of the state change. 

\emph{iii)} The \textit{Concrete Test Case Generation} activity uses the augmented GUI model to identify the sequences of UI interactions that exercise the test sequences identified in the first step as the ones relevant to verify the behavior of the enforcer. These UI interactions, enriched with program oracles, are the test cases that can be executed to validate the activity of the enforcer. In the following, we briefly describe each activity, explaining how we extended the approach to collect and verify the non-functional impact of the enforcer. More information about Test4Enforcers can be found in~\cite{Guzman:Test4Enforcer:RV:2020}.

\textbf{\textit{Generation of the Test Sequences}}. In this activity, Test4Enforcers generates the sequences of operations that must be exercised to validate an enforcer whose behavior is captured by an enforcement model \textcolor{\rev}{(we assume the model is correctly specified and verified \cite{riganelli2017verifying})}. To this end, Test4Enforcers uses the \emph{Harmonized State Identifiers} (HSI) method~\cite{luo1995selecting,belli2015fault}, which is a variant of the W-method~\cite{chow1978testing,lee1996principles,sidhu1989formal} that does not require the model of the system to be completely specified. In fact, not every combination of event is feasible in an enforcement model, for instance due to the constraints of the environment (e.g., it is not possible to pause an app that is already paused).

We use HSI since it works well to reveal implementation errors, such as erroneous next-states, extra/missing states, etc. In a nutshell, HSI supports the generation of tests that cover every transition in the model and that check the identity of the state reached by each transition to ensure that the implementation of the model is correct.

HSI exploits the notions of \textit{transition cover}, to cover all the transitions, and \textit{separating families}, to check the identity of the states reached by each transition. Informally, a transition cover $P$  for an automaton $A$ is a set of input sequences, including the empty sequence $\epsilon$, that exercises all the transitions in A (every transition must occur as the last element of at least an input sequence). 
A separating family $H$ includes an element $H_i$ for each state $s_i$ of the mode. The element $H_i$ is a set input sequences that can be executed to distinguish the state $s_i$ from the other states of the system (that is, the outputs observed by executing the sequences of inputs in $H_i$ from $s_i$ are different from the ones produced by any other state of the system). 

The final set of test sequences are obtained by concatenating each element $tc_i$ in the transition coverage set $P$ with every input sequence in the separating family associated with the state reached after executing $tc_i$. Namely, if $tc_i$ reaches state $s_i$, the elements in the separating family $H_i$ are concatenated to $tc_i$ to obtain the final set of test sequences (note that the prefix $t_i$ must be executed multiple times, depending on the number of elements in $H_i$). This process may generate redundant combinations that are filtered out from the final set of input sequences to be exercised.

In our example, this process generates the following sequences to be covered with test cases:\\


\noindent \quad $activity.onPause()_{req}$, 

\noindent \quad $activity.onPause()_{req}\ activity.onPause()_{req}$,

\noindent \quad $camera.open()_{req}\ activity.onPause()_{req}$, 

\noindent \quad $camera.open()_{req}\ activity.onPause()_{req}\ activity.onPause()_{req}$, 

\noindent \quad $camera.open()_{req}\ camera.release()_{req}\ activity.onPause()_{req}$ \\

\textbf{\textit{GUI Ripping with Tracing}}. GUI Ripping is an exploration strategy that can be used to explore the GUI of an app under test with the purpose of building a state-based representation of its behavior~\cite{Memon:Ripping:WCRE:2013}. GUI ripping generates the state-based model of the app under test by systematically executing every possible action on every state encountered during the exploration, until a given time or action budget expires. Our implementation of Test4Enforcers targets Android apps and uses DroidBot~\cite{Droidbot} configured to execute actions in a breadth-first manner to build the state-based model. 

A state $s$ of the app under test is represented by its visible views $s=\{v_i | i=1 \ldots n\}$. Each view $v_i$ is defined by a set of properties $v_i=\{p_{i1},\ldots, p_{ik}\}$, with each property being a key-value pair.  For instance, \texttt{EditText} is an Android view for entering text and it has properties such as \texttt{clickable}, to specify if the view reacts to click events, and \texttt{text} to store the text present in the view. 

Operations that change the set of visible views (e.g., because an activity is closed and another one is opened) or the properties of the views (e.g., because some text is entered in an input field) change the state of the app. DroidBot uses the following set of actions $A_{app}$ during GUI ripping: \emph{touch} and \emph{long touch}, which execute a tap and a long tap on a clickable view, respectively; \emph{setText}, which enters a pre-defined text inside an editable view; \emph{keyEvent}, which presses a navigation button; and \emph{scroll}, which scrolls the current window. 

The state-based representation of the execution space produced by GUI Ripping includes all the visited states and the executed actions. Test4Enforcers enriches the model generated by GUI ripping with the information reported by a tracer, which associates each transition with a sequence of methods belonging to the enforcer's alphabet, if executed during the transition between state. The state-based model thus shows both the UI interactions that can be executed on the app, their effect on the state of the app, and the internal events that are activated when they are executed. 

\begin{figure}[]
\centering
        \includegraphics[width=\textwidth]{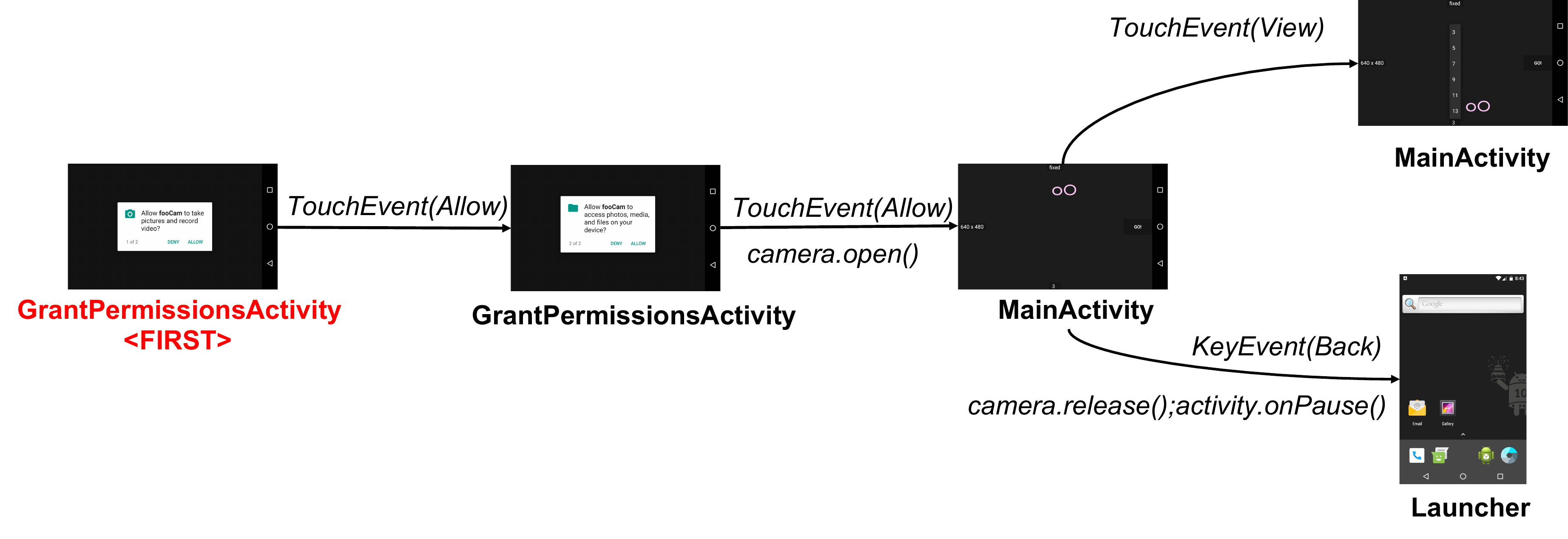}
\caption{Excerpt of a model derived with GUI Ripping.}\label{fig:guimodel}
\end{figure}

Figure~\ref{fig:guimodel} shows an excerpt of the model obtained by running the ripping activity on the \texttt{fooCam} app~\footnote{fooCam is a HDR camera app that can take multiple shots with different exposure settings. The app is available on the Google Play Store at the following link: \url{https://play.google.com/store/apps/details?id=net.phunehehe.foocam2&hl=EN}.} while considering the alphabet of the enforcer shown in Figure~\ref{fig:enforcer-composition}. For simplicity, we represent the states with the screenshots of the app. The labels above transitions represent UI interactions, while the labels below transitions, when present, represent events in the alphabet of the enforcer collected by the monitor. For instance, when the \emph{KeyEvent(Back)} UI interaction is executed and the app moves from the state \emph{MainActivity} to the state \emph{Launcher}, the sequence of internal events \emph{camera.release()} \emph{activity.onPause()} is observed.  

\textbf{\textit{Concrete Test Case Generation}}. Generating concrete (i.e., executable) test cases consists of finding the sequences of GUI interactions that cause the execution of the desired sequences of events belonging to the alphabet of the enforcer, as identified by the HSI method. To this end, Test4Enforcers exploits both the augmented model derived with GUI ripping and the sequences generated with the HSI method. In particular, Test4Enforcers aims to generate a test suite that contains a test case for each target sequence of events generated with the HSI method. These tests are identified by searching for a path in the GUI model that covers the sequence under consideration according to the annotations on the transitions. For instance, if the sequence to be covered is \textit{camera.open()} \textit{camera.release()} \textit{activity.onPause()} and the GUI model is the one in Figure~\ref{fig:guimodel}, the sequence \textit{TouchEvent(Allow)} \textit{TouchEvent(Allow)} \textit{KeyEvent(Back)} is identified as the concrete test to execute. In fact, the execution of the identified UI events is expected to produce the desired internal computation (based on the labels on the transitions). 

\textcolor{\rev}{If there is at least one path in the GUI model that covers the sequence under consideration, the corresponding test case is generated. Since a path derived from the model is not necessarily feasible,} Test4Enforcers identifies the 10 shortest paths that cover the target sequence and executes them sequentially until it is able to exercise the target sequence. If the right test is found, a \emph{differential oracle} is embedded in the test case. A differential oracle is an oracle that determines the correctness of a test execution by comparing two executions of the same test on two different programs. In our case, the compared programs are the app \emph{with} and \emph{without} the enforcer deployed. Test4Enforcers injects two different differential oracles: the \emph{transparent-enforcement oracle} and the \emph{actual-enforcement oracle}. 


A test is assigned with the \emph{transparent-enforcement oracle} when the test must \emph{produce the same result if executed on both the apps with and without the enforcer in place}. In other words, the exercised sequence does not require any change performed by the enforcer. For instance, the sequence \textit{camera.open()} \textit{camera.release()} \textit{activity.onPause()} is not altered by the enforcer in Figure~\ref{fig:enforcer-composition} and thus the transparent-enforcement oracle is used to determine the correctness of the test that covers this sequence, that is, no behavioral differences must be observed when this sequence is executed in both the app without and the app with the enforcer. 


A test is assigned with the \emph{actual-enforcement oracle} when the test must \emph{produce a different outcome when executed on the app with and without the enforcer in place}. In other words, the tested sequence corresponds to a path that requires the intervention of the enforcer. For instance, the sequence \textit{camera.open()} \textit{activity.onPause()} causes the intervention of the enforcer shown in Figure~\ref{fig:enforcer-composition}, which outputs the extra event \textit{camera.release()}. The test corresponding to that sequence is thus labeled as \emph{producing the same result until the \textit{activity.onPause()} event, and a potentially different result afterwards}, and the actual-enforcement oracle is embedded in the test.



\subsection{Test Execution}
The test  execution step aims to automatically identify, if present, any problem introduced by the enforcer in the target app. The transparent-enforcement and actual-enforcement oracles can detect functional misbehaviors. Here we describe how Test4Enforcers can detect non-functional problems, performance degradations in particular, introduced by enforcers.

Test execution requires three inputs: $1)$ The enforcer implementation that must be tested; $2)$ One or more apps to be used to concretely test the enforcer; and $3)$ The automatic test suites generated in the previous step to exercise the enforcer's behavior.  

\begin{figure}
    \centering
        \includegraphics[width=12cm]{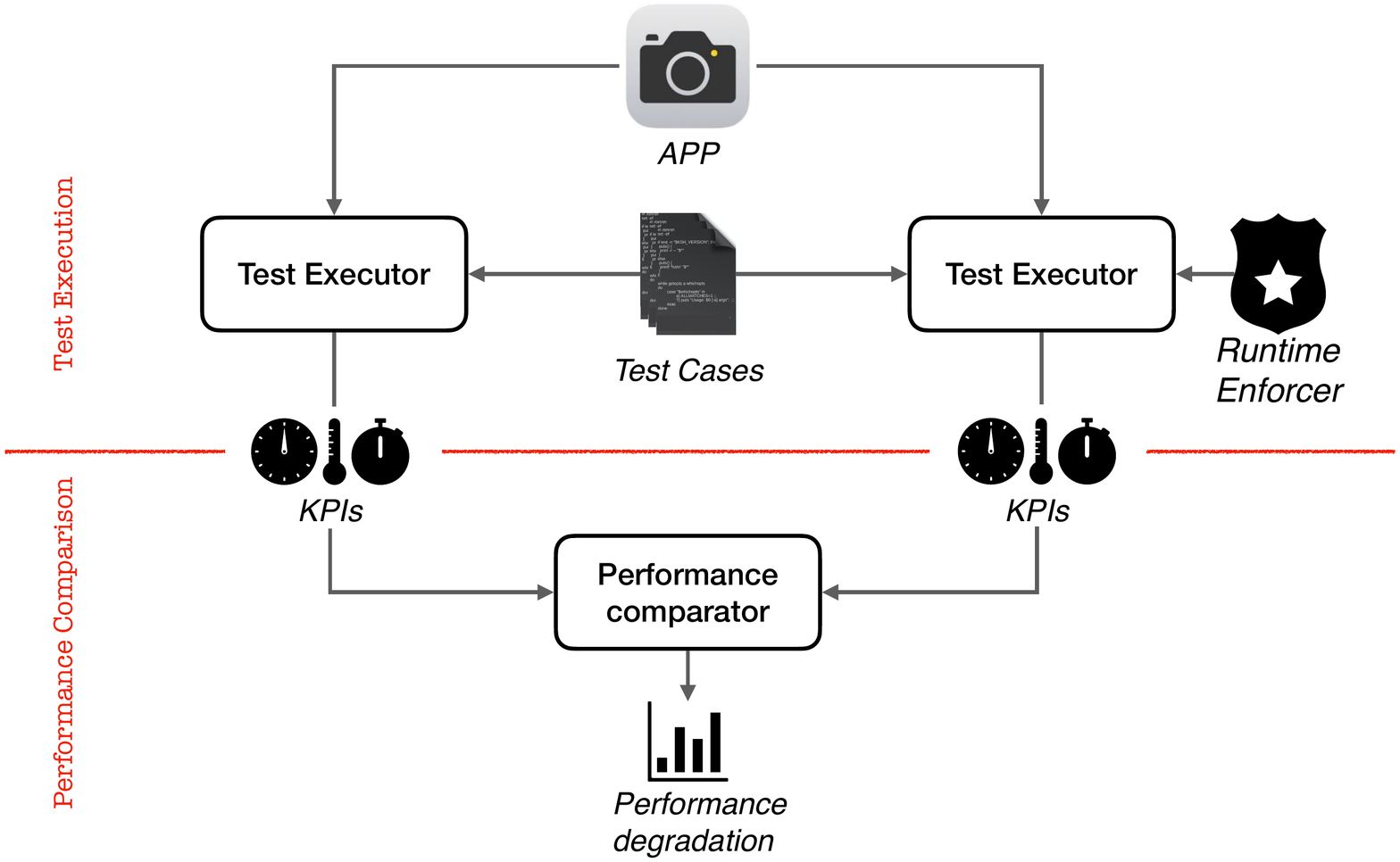}
    \caption{Detection of Non-Functional Programs with Test4Enforcers.}
    \label{fig:TCExecution}
\end{figure}

The output is a list of KPIs whose values indicate performance degradation when the enforcer is active. As shown in Figure~\ref{fig:TCExecution}, this step performs two main activities: \emph{i)} \emph{Test Execution}, which runs the test suite that exercises the enforcer behaviour and collects the KPI values from executions with and without the enforcer, and \emph{ii)} \emph{Performance Comparison}, which identifies performance degradation by comparing the KPI values obtained with and without the enforcer. 

\textbf{\textit{Test Execution}}. This activity runs the automatic test suite and collects KPI values about the performance of the running app. Specifically, we used profiler tools~\cite{dumpsys,AndroidProfiler} that come with Android Studio \cite{AndroidStudio} and allow us to obtain KPIs on CPU, memory, network, and energy usage.  The test suite is executed ten times to compute stable median values not affected by outliers for each KPI.

The execution of the test suite is performed by two instances of the Test Executor components, which can be executed either sequentially or in parallel (e.g., on a cloud infrastructure). One instance runs the test suite on the unaltered version of the app under test, to collect KPIs about the app's performance when no enforcer is deployed. The collected values represent the baseline for the detection of performance degradation. The other instance of the Test Executor runs the test suite on the app enriched with the software enforcer, to collect data about the potential degradation of the performance. 


\textbf{\textit{Performance Comparison}}. 
The performance comparison activity interprets the results produced by Test Execution and reports any performance degradation. We currently support four non-functional performance characteristics \cite{SurveyPerformanceOptimizationAndroid:2021}: responsiveness, launch time, memory consumption, and energy consumption. \textcolor{\rev}{The reported performance degradations depend on some thresholds and parameters whose default values can be changed by users according to the specific context. We report below the values we used in our evaluation.} 

 \textit{Responsiveness} refers to the ability of a mobile app to respond to user iterations in a timely manner. A highly responsive app improvers the user satisfaction, since users prefer to not wait for too long when interacting with an app. On the other hand, an app with poor responsiveness can have a negative impact on user experience and its success on the market. Consequently, the enforcer should not negatively impact responsiveness.

In the case of Android apps, response times of less than 200 ms is known to not negatively impact user satisfaction \cite{TestingAndroidResponsiveness:2013}, while delays that last almost a second do not significantly affect users but can be recognized by users, while longer delays have a clearly negative impact on the user experience.

By measuring the CPU usage time in event handler methods, we can measure the response time to GUI events. If any UI interaction requires more than 200 ms~\cite{TestingAndroidResponsiveness:2013} to be served only when the enforcer is active, a poor responsiveness caused by the enforcer is reported to the user. 

 \textit{Launch time} is another KPI analyzed by Test4Enforcers. It represents the amount of time necessary to boot the app. During start-up, the app initializes all its components, including initializing objects, creating and initializing activities, inflating the layout, and drawing the app for the first time. This is the first performance characteristic that a user is exposed to. Since from the first usage, users normally expect a short launch time, otherwise, the experience and satisfaction might be negative. The launch time is considered to be excessive when it exceeds 5 seconds, as reported in the Android documentation \cite{AppStartupTime}. 
 
To detect any degradation of this performance  characteristic, Test4Enforcers measures the time the app needs to produce the first frame. If the value of this KPI exceeds 5 seconds only when the enforcer is active then a degradation of the startup time caused by the enforcer is reported to developers. 
 
\textit{Memory consumption} is another important performance characteristic for mobile apps as these are running on devices with limited resources. For instance, a memory leakage can lead to performance degradation due to the frequent triggering of the garbage collection process. These slowdowns can then evolve into freezes and even app crashes. In order to be able to detect a degradation in memory consumption, we analyze KPIs that can give information regarding the memory usage and report a degradation when the execution with the enforcer introduces an overhead of more than 5\% compared to the baseline without enforcer . 

\textit{Power consumption} is a crucial characteristic of mobile systems. Running an enforcer with abnormal power consumption can drastically reduce battery life, affecting the user experience. 
Therefore, it is very important to solve problems of energy inefficiencies introduced by enforcers.
To detect energy hotspots and bugs introduced by enforcers, Test4Enforcers collects and analyzes KPI values regarding battery usage, and reports a degradation when execution with the enforcer introduces more than 5\% overhead compared to baseline without enforcers. 

%% file: evaluation.tex
\section{Evaluation} \label{sec:evaluation}
We evaluated the capability of Test4Enforcers to capture performance degradations introduced by faulty enforcers. In particular, we addressed the following research question: \emph{Can Test4Enforcers detect actual performance bugs?}

To investigate this research question, we selected fooCam as the app under test. It is a camera app that can automatically take multiple successive shots with different exposure settings. FooCam has been already used in previous works about API misuses \cite{Liu:FixingResourceLeaks:ISSRE:2016,riganelli2017policy,Riganelli:ProactiveLibraries:ACMTAAS:2019}. 

The enforcer used to validate the effectiveness of Test4Enforcers in detecting performance bugs is an implementation of the enforcement model shown in Figure \ref{fig:enforcer-composition} that has been tested to have no functional bugs \cite{Guzman:Test4Enforcer:RV:2020}.

To evaluate the performance of Test4Enforces, we injected bugs that can cause performance degradation into the source code of the enforcer.  In particular, we injected 4 bugs  that affect the main non-functional performance characteristics of mobile apps \cite{SurveyPerformanceOptimizationAndroid:2021}:
\begin{itemize}
\item A responsiveness bug that introduces a delay in reacting to GUI events.
\item An app startup bug that delays the loading of the app's initial screen.
\item A CPU hog bug that executes some CPU intensive tasks that significantly increase memory consumption
\item A memory leak bug that executes tasks that allocate memory without releasing it.
\end{itemize}

     \begin{figure}
        \centering
        \begin{subfigure}[b]{0.47\textwidth}
            \centering
            \includegraphics[width=\textwidth]{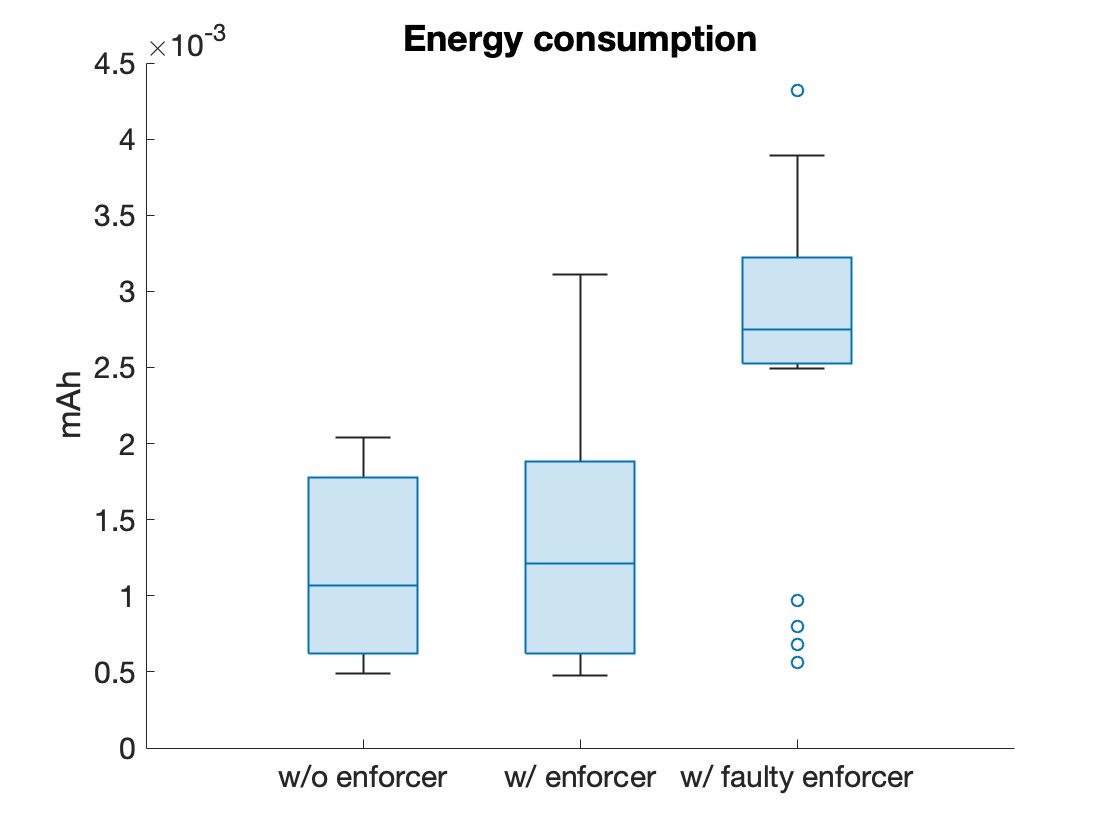}
            \caption[estimated power use]%
            {{\small Estimated Power Use}}    
            \label{fig:energymetric}
        \end{subfigure}
        \hfill
        \begin{subfigure}[b]{0.47\textwidth}  
            \centering 
            \includegraphics[width=\textwidth]{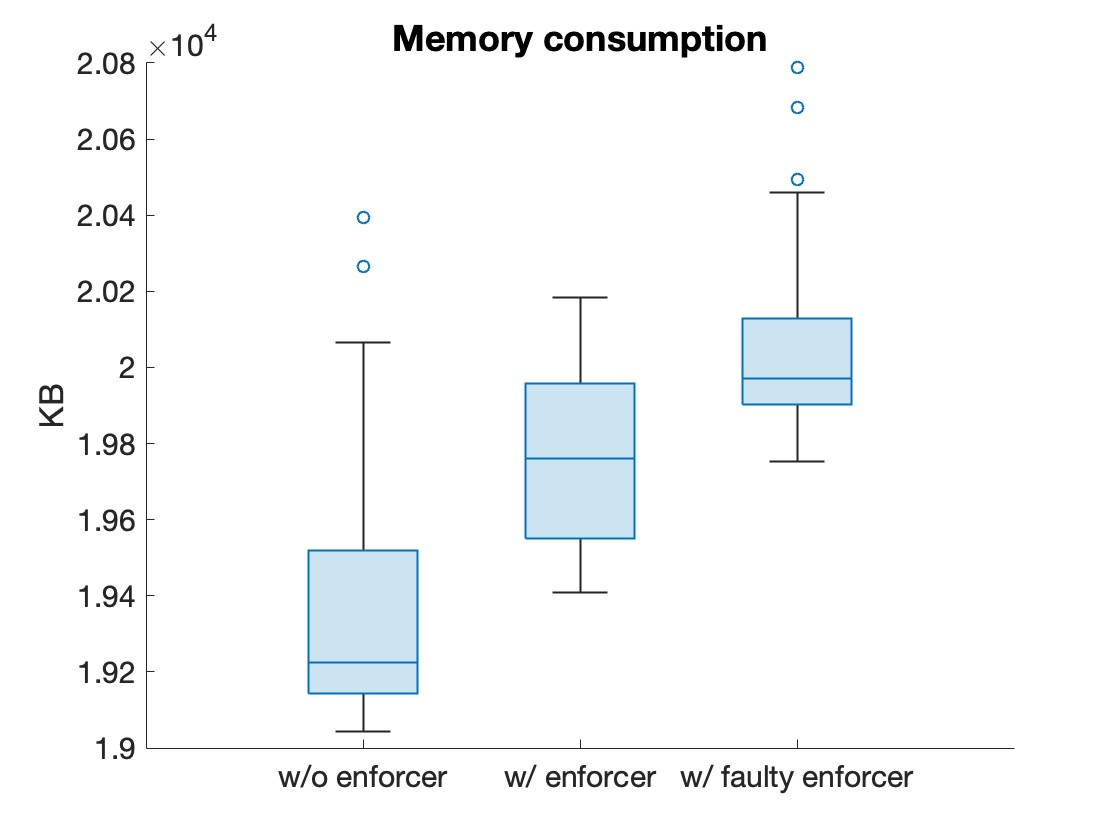}
            \caption[Total Proportional Set Size]%
            {{\small Total Proportional Set Size}}    
            \label{fig:memorymetric}
        \end{subfigure}
        \vskip\baselineskip
        \begin{subfigure}[b]{0.47\textwidth}   
            \centering 
            \includegraphics[width=\textwidth]{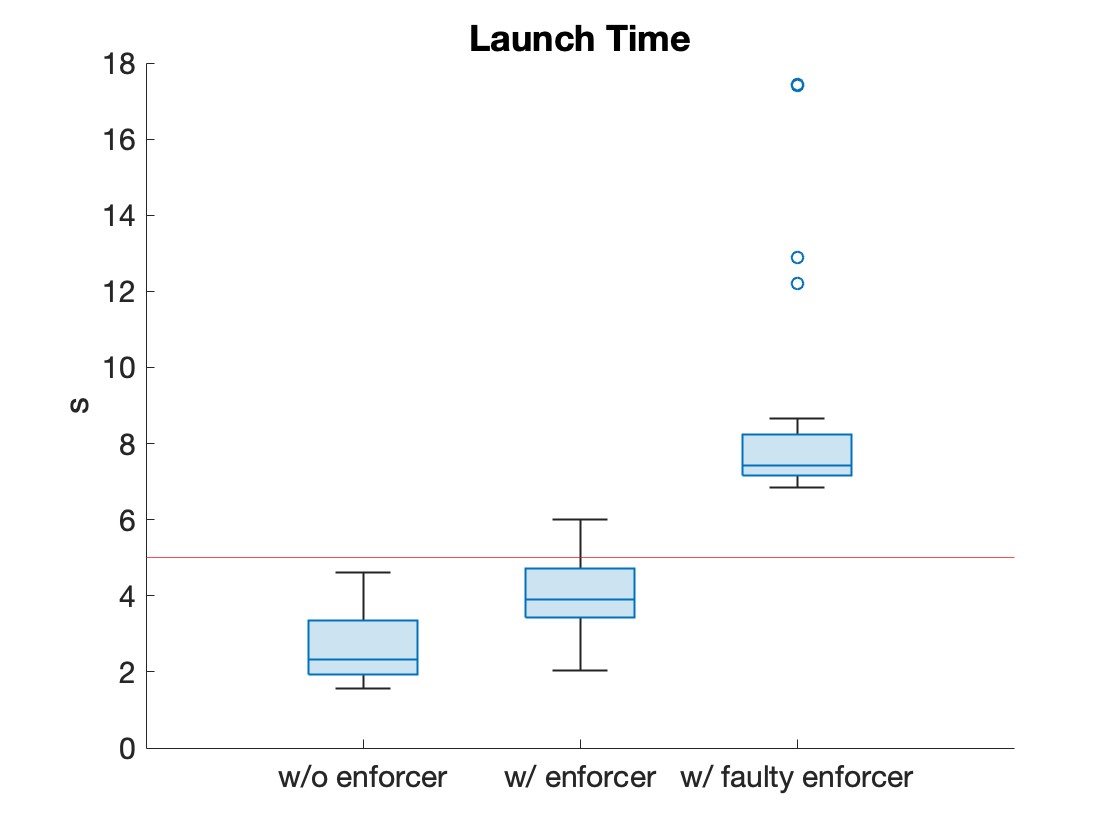}
            \caption[]%
            {{\small Time to Initial Display}}    
            \label{fig:launchtimemetric}
        \end{subfigure}
        \hfill
        \begin{subfigure}[b]{0.47\textwidth}   
            \centering 
            \includegraphics[width=\textwidth]{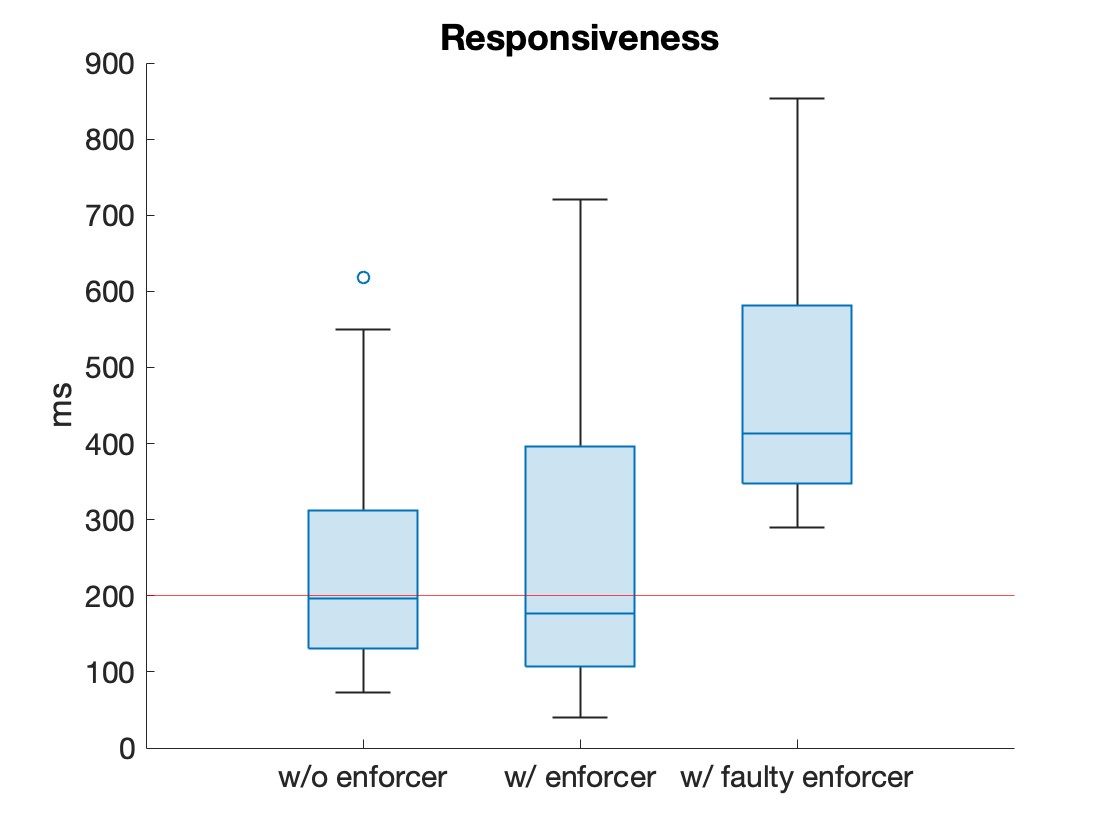}
            \caption[]%
            {{\small GUI Event Handler Execution Time}}    
            \label{fig:responsivenessmetric}
        \end{subfigure}
        \caption[ Comparison of KPIs in executions without enforcer, with enforcer, and with buggy enforcer.]
        {\small Comparison of KPIs in executions without enforcer, with enforcer, and with faulty enforcer.} 
        \label{fig:metrics}
    \end{figure}

Figure~\ref{fig:metrics} shows the boxplots that compare the collected values of the KPIs, one for each performance characteristic validated by Test4Enforcers. The three boxes in each plot shows the distribution of values obtained when the app is executed without the enforcer (label \textit{w/o enforcer}), with the correct enforcer  (label \textit{with enforcer}) and with the faulty enforcer (label \textit{with faulty enforcer}). \textcolor{\rev}{Each box represents a population of 10 samples obtained by running the same test case 10 times.}

Figure~\ref{fig:energymetric} shows the values obtained for the \textit{Estimated Power Use} KPI, which calculates the app's power consumption. \textcolor{\rev}{We can notice that the enforcer has a small impact on energy consumption, which confirms the suitability of the enforcement technology on the power perspective.} The faulty enforcer instead introduces the battery usage by more than 14\%, which can be detected by Test4Enforcers when comparing the collected KPIs.

Figure~\ref{fig:memorymetric}  shows the values of the KPI ``Total Propotional Set Size'' (PSS), which represents the RAM used by the app. Test4Enforcers can detect the problem introduced by the faulty enforcer since memory consumption is increased by more than 5\%, while it is not the case when the correct enforcer is introduced. In this case, we could notice a difference in memory consumption since the enforcer has to collect and store information about the status of the system. However, the memory overhead is below 5\%.

Figures~\ref{fig:launchtimemetric} and~\ref{fig:responsivenessmetric} report the collected values for launch time and responsiveness. The red lines in these plots represent the threshold values (5 seconds and 200 milliseconds) that should not be passed, as reported in the literature and discussed in the previous section. Results show that the correct enforcers introduce a marginal overhead in launch time and responsiveness, while the faulty enforcers introduce a performance degradation that could be automatically detected.


In conclusion, Test4enforcers was always able to stimulate the behavior of the enforcers and correctly evaluates their performance, helping developers timely identifying performance bugs.

%% file: RW.tex
   
\section{Related Work}\label{sec:related}

\emph{Runtime enforcement} is a powerful technique that enforces the behaviour specified by a set of correctness policies on a running system. In particular, runtime enforcement strategies modify executions assuring that policies are satisfied despite the potentially incorrect behavior of the monitored software~\cite{Barringer2010,survey2012}. 
Runtime enforcement has been applied in multiple domains, including mobile apps~\cite{Riganelli:HealingDataLos:IWSF:2016,riganelli2017policy,Falcone:AndoridEnforcement:RV:2012,DaianFMSSIR15} operating systems~\cite{Sidiroglou_AAS_2009}, web-based applications~\cite{Magalhaes_SSH_2015}, control systems~\cite{ControlSystem:RE:Lanotte:2020}, and cloud systems~\cite{Dai_SHD_2009}. Among these many domains, in this paper we focus on the Android environment, which has been already considered in the work by Falcone et al.~\cite{Falcone:AndoridEnforcement:RV:2012}, who studied how to enforce privacy policies by detecting and disabling suspicious method calls, and more recently by Riganelli et al.~\cite{riganelli2017policy,Riganelli:ProactiveLibraries:ACMTAAS:2019,Riganelli:EnforcerReusability:ISOLA:2018}, who studied how to augment classic Android libraries with proactive mechanisms that can automatically suppress and insert API calls to enforce resource usage policies. 





While runtime enforcement strategies focus on the definition of models and strategies to specify and implement the enforcers, Test4Enforcers is complemental to this effort, since it derives the test cases that should be executed on apps with and without the enforcers to verify the correctness and performance of the implemented enforcer. In particular, performance is a crucial issue especially in mobile systems that have limited resources.



\smallskip

\emph{Performance testing} refers to the execution of test cases to evaluate the behavior of an application with respect to performance aspects, such as responsiveness or efficiency \cite{molyneaux2014art}. Despite the large number of approches aimed at automating functional testing in mobile apps \cite{TestingAndroidApps:2019}, studies highlight the lack of approaches that include performance testing \cite{ChallengesMobileTestingMuccini:2012,AutomatedMobileAppTesting:2017}. Developers depend on manual testing and analysis to detect performance bugs \cite{linares2015developers} by using tools for profiling and debugging their applications \cite{SurveyPerformanceOptimizationAndroid:2021,profileApp}.  Test4Enforcers automates the performance testing of runtime enforcers in mobile devices by automatically generating and executing test cases that interact with an app's GUI  in order to verify both the behavior and performance of an enforcer.

\smallskip
\emph{Verification and Validation of runtime enforcement} concerns with checking that the software enforcer is indeed delivering the intended behavior. In fact, although the enforcer is meant to correct the behavior of a monitored software, the enforcer itself might still be wrong and its activity might compromise the correctness of the system rather than improving it.  
A recent work in this direction is the one by Riganelli et al.~\cite{riganelli2017verifying} that provides a way to verify if the activity of multiple enforcers may interfere. The proposed analysis is however entirely based on the models and the many problems that might be introduced by the actual software enforcers cannot be revealed with that approach. Test4Enforcers provides a complemental capability, that is, it can test if the implementation of the enforcer behaves as expected once injected in the target system.
